\documentclass[pra]{revtex4}
\usepackage{amsmath}
\usepackage{graphics}
\usepackage{epsfig}
\usepackage{xspace}
\usepackage{soul}
\usepackage{amssymb}
\usepackage{color}
\usepackage{subfigure}

\begin{document}

\title{Making the $\mathcal{PT}$ symmetry unbreakable }
\author{ Vitaly Lutsky$^{1}$, Eitam Luz$^{1}$, Er'el Granot$^{2}$, and Boris
A. Malomed$^{1,3,4}$}
\affiliation{$^{1}$Department of Physical Electronics, School of Electrical Engineering,
Faculty of Engineering, Tel Aviv University, Tel Aviv 69978, Israel\\
$^{2}$Department of Electrical and Electronic Engineering, Ariel University,
Ariel, Israel \\
$^{3}$Center for Light-Matter Interaction, Tel Aviv University, Tel Aviv
69978, Israel\\
$^{4}$ITMO University, St.Petersburg 197101, Russia}

\begin{abstract}
It is well known that typical $\mathcal{PT}$-symmetric systems suffer
symmetry breaking when the strength of the gain-loss terms, i.e., the
coefficient in front of the non-Hermitian part of the underlying
Hamiltonian, exceeds a certain critical value. In this article, we present a
summary of recently published and newly produced results which demonstrate
various possibilities of extending the $\mathcal{PT}$ symmetry to
arbitrarily large values of the gain-loss coefficient. First, we
recapitulate the analysis which demonstrates a possibility of the
restoration of the $\mathcal{PT}$ symmetry and, moreover, complete avoidance
of the breaking in a photonic waveguiding channel of a subwavelength width.
The analysis is necessarily based on the system of Maxwell's equations,
instead of the usual paraxial approximation. Full elimination of the $%
\mathcal{PT}$-symmetry-breaking transition is found in a deeply
subwavelength region. Next, we review a recently proposed possibility to
construct stable one-dimensional (1D) $\mathcal{PT}$-symmetric solitons in a
paraxial model with arbitrarily large values of the gain-loss coefficient,
provided that the self-trapping of the solitons is induced by
self-defocusing cubic nonlinearity, whose local strength grows sufficiently
fast from the center to periphery.\ The model admits a particular analytical
solution for the fundamental soliton, and provides full stability for
families of fundamental and dipole solitons. It is relevant to stress that
this model is nonlinearizable, hence the concept of the $\mathcal{PT}$
symmetry in it is also an essentially nonlinear one. Finally, we report new
results for unbreakable $\mathcal{PT}$-symmetric solitons in 2D extensions
of the 1D model: one with a quasi-1D modulation profile of the local
gain-loss coefficient, and another with the fully-2D modulation. These
settings admit particular analytical solutions for 2D solitons, while
generic soliton families are found in a numerical form. The quasi-1D
modulation profile gives rise to a stable family of single-peak 2D solitons,
while their dual-peak counterparts tend to be unstable. The soliton
stability in the full 2D model is possible if the local gain-loss term is
subject to spatial confinement.
\end{abstract}

\maketitle


\section{Introduction}

A fundamental principle of the quantum theory is that, while the underlying
wave function may be complex, eigenvalues of energy and other physically
relevant quantities must be real, which is provided by the condition that
the respective Hamiltonian is self-conjugate (Hermitian) \cite{qm}. On the
other hand, the condition of the reality of the entire energy spectrum does
not necessarily imply that it is generated by a Hermitian Hamiltonian.
Indeed, it had been demonstrated, about twenty years ago, that non-Hermitian
Hamiltonians obeying the parity-time ($\mathcal{PT}$) symmetry may also
produce entirely real spectra~\cite%
{bender1,dorey,bender2,bender3,bender4,review,ptqm}. In terms of the usual
single-particle Hamiltonian, which includes potential $U(\mathbf{r})$, the $%
\mathcal{PT}$ symmetry implies that the potential is complex, $U(\mathbf{r}%
)=V(\mathbf{r})+iW(\mathbf{r})$ (the usual Hermitian Hamiltonian contains a
strictly real potential), its real and imaginary parts being, respectively,
even and odd functions of coordinates \cite{bender1}:.
\begin{equation}
V(\mathbf{r})=V(-\mathbf{r}),W(-\mathbf{r})=-W(\mathbf{r}),~\mathrm{i.e.,~}%
U(-\mathbf{r})=U^{\ast }(\mathbf{r}),  \label{minus}
\end{equation}%
~ where $\ast $ stands for the complex conjugate. For a given real part of
the potential, the spectrum of $\mathcal{PT}$-symmetric models remains
completely real, i.e., physically relevant, as long as the strength of the
imaginary component of the potential is kept below a certain critical value,
which is a threshold of the $\mathcal{PT}$-symmetry breaking, above which
the system becomes unstable. The loss of the $\mathcal{PT}$ symmetry may be
preceding by the onset of the jamming anomaly, which means transition from
increase to decreases of the power flux between the gain and loss elements
in the system following the increase of the gain-loss coefficient \cite%
{jamming1,jamming2}. It is relevant to mention that some relatively simple $%
\mathcal{PT}$-symmetric systems may be explicitly transformed into an
alternative form admitting a representation in terms of an Hermitian \
Hamiltonian \cite{Barash2,Barash}.

While the concept of $\mathcal{PT}$-symmetric Hamiltonians remained an
abstract one in the framework of the quantum theory per se, theoretical
works had predicted a possibility to emulate this concept in optical media
with symmetrically placed gain and loss elements \cite{theo1}-\cite{Kominis}%
, making use of the commonly known similarity between the Schr\"{o}dinger
equation in quantum mechanics and the classical equation governing the
paraxial light propagation in classical waveguides. These predictions were
followed by the implementation in optical waveguiding settings of various
types \cite{exp1}-\cite{exp7}, as well as in metamaterials \cite{exp4},
lasers \cite{exp5} (and laser absorbers \cite{Longhi}), microcavities \cite%
{exp6}, optically induced atomic lattices \cite{exp8}, exciton-polariton
condensates \cite{exci1}-\cite{exci3}, and in other physically relevant
contexts. In particular, the transitions from unbroken to broken $\mathcal{PT%
}$ symmetry was observed in many experiments. One of prominent
experimentally demonstrated applications of the $\mathcal{PT}$ symmetry in
optics is unidirectional transmission of light \cite{uni}.

Other classical waveguiding settings also admit emulation of the $\mathcal{PT%
}$ symmetry, as demonstrated in acoustics \cite{acoustics} and predicted in
optomechanical systems \cite{om}. Also predicted were realizations of this
symmetry in atomic Bose-Einstein condensates \cite{Cartarius}, magnetism
\cite{magnetism}, mechanical chains of coupled pendula \cite{Peli}, and
electronic circuits \cite{electronics} (in the latter case, the prediction
was also demonstrated experimentally). In terms of the theoretical analysis,
$\mathcal{PT}$-symmetric extensions were also elaborated for Korteweg - de
Vries \cite{KdV1,KdV2}, Burgers \cite{Zhenya-Burgers}, and sine-Gordon \cite%
{Cuevas} equations, as well as in a system combining the $\mathcal{PT}$
symmetry with the optical emulation of the spin-orbit coupling \cite{HS}.

While the $\mathcal{PT}$ symmetry is a linear property of the system, it may
be naturally combined with intrinsic nonlinearity of the medium in which the
symmetry is realized, such as the ubiquitous Kerr nonlinearity of optical
waveguides. Most typically, these settings are modelled by nonlinear Schr%
\"{o}dinger equations (NLSEs) with the $\mathcal{PT}$-symmetric potentials,
subject to constraint (\ref{minus}), and cubic terms. Such models may give
rise to $\mathcal{PT}$-symmetric solitons, which were considered, chiefly
theoretically, in a large number of works (see, in particular, theoretical
papers \cite{soliton}, \cite{Konotop}-\cite{Alexeeva} and recent reviews
\cite{review1,review2}), and experimentally demonstrated too \cite{exp7}.
While most of these works were dealing with one-dimensional (1D) models,
stable $\mathcal{PT}$-symmetric solitons were also found in some
two-dimensional (2D) models \cite{Yang}, \cite{2D-1}-\cite{2D-4}, \cite{HS}.
A characteristic feature of solitons in $\mathcal{PT}$-symmetric systems is
that, although these systems model, generally speaking, dissipative dynamics
(the systems have no dynamical invariants), their solitons form continuous
families like in conservative systems (defined by usual Hermitian
Hamiltonians) \cite{families}, while traditional dissipative nonlinear
systems normally give rise to isolated solutions in the form of dissipative
solitons, which do not form families (if a dissipative soliton is stable, it
plays the role of an attractor in the system's dynamics \cite{diss1}-\cite%
{diss3}).

Similar to their linear counterparts, soliton states are also subject to
destabilization via the breaking of the $\mathcal{PT}$ symmetry at a
critical value of the strength of the gain-loss terms \cite{breaking}.
Nevertheless, there are specific models which make the solitons' $\mathcal{PT%
}$ symmetry \emph{unbreakable}, extending it to arbitrarily large values of
the gain-loss strength, i.e., the coefficient in front of the non-Hermitian
part of the respective Hamiltonian \cite{unbreakable}-\cite{China}. The
particular property of those models is that self-trapping of solitons is
provided not by the usual self-focusing sign of the cubic nonlinearity, but
by the opposite defocusing sign, with the local strength of the
self-defocusing growing fast enough from the center to periphery. For
conservative systems (in the absence of gain and loss), this scheme of the
self-trapping of stable 1D, 2D, and 3D solitons was elaborated previously in
a number of works \cite{Barcelona0}-\cite{Barcelona11}.

The objective of the present article is to provide a brief survey of systems
which may support unbreakable $\mathcal{PT}$ symmetry, as this property is
quite promising for potential applications, and is interesting in its own
right. It was recently elaborated in two completely different settings. One
is the above-mentioned model with the solitons supported by the spatially
growing strength of local self-defocusing. On the other hand, a possibility
of creating the $\mathcal{PT}$ symmetry persisting up to indefinitely large
values of the gain-loss coefficient was also discovered in the context of
nanophotonics, considering light propagation in structures combining
refractive, amplifying, and attenuating elements at a subwavelength scale
\cite{sub}. This setting was theoretically analyzed in a purely linear form,
with an essential peculiarity that the corresponding model is, naturally,
based on the full system of Maxwell's equations, rather than the
paraxial-propagation equation of the Schr\"{o}dinger type, which was used in
an absolute majority of works dealing with the $\mathcal{PT}$ symmetry in
optical media. Basic findings for the restoration of the $\mathcal{PT}$
symmetry, and a possibility of making it completely unbreakable in the
linear nanophotonic model are presented below in Section II.

The results for unbreakable 1D $\mathcal{PT}$-symmetric solitons in the
model based on the paraxial-propagation NLSE with the spatially growing
strength of the self-defocusing nonlinearity are summarized in Section III.
It is followed by Section IV, which reports \emph{new results} for 2D
extensions of the unbreakable $\mathcal{PT}$ symmetry in a nonlinear model
of a similar type. We consider two different versions of the 2D system, with
the quasi-one-dimensional or fully two-dimensional $\mathcal{PT}$ symmetry,
the former meaning that the gain and loss are swapped by reflection $%
x\Leftrightarrow -x$, while the reflection in the perpendicular direction, $%
y\Leftrightarrow -y$ leaves the gain-loss pattern invariant. The main issue
is the stability of the 2D $\mathcal{PT}$-symmetric solitons, which turn out
to be essentially more stable in the case of the quasi-1D symmetry than in
the framework of the full 2D scheme. An essential asset of the 1D, quasi-1D,
and full 2D models is that a number of soliton solutions can be obtained in
an exact analytical form, even if not all of them are stable.

\section{Restoration and persistence of the $\mathcal{PT}$ symmetry in the
photonic medium with a subwavelength structure}

Following Ref. \cite{sub}, we here consider the propagation of monochromatic
light beams with the TM (transverse-magnetic) polarization , which include
only $\mathcal{E}_{x}$, $\mathcal{E}_{z}$, and $\mathcal{H}_{y}$ components
of the electric and magnetic fields. The propagation is considered along the
$z$ axis in an effectively 2D medium whose dielectric permittivity is
modulated in the transverse direction, $x$. The spatial evolution of the
field components is governed by the reduced system of the Maxwell's
equations:%
\begin{gather}
i\frac{\partial E_{x}}{\partial z}=-\frac{1}{\varepsilon _{0}\omega }\frac{%
\partial }{\partial x}\left( \frac{1}{\varepsilon _{\mathrm{rel}}}\frac{%
\partial \mathcal{H}_{y}}{\partial x}\right) -\mu _{0}\omega \mathcal{H}_{y},
\notag \\
i\frac{\partial \mathcal{H}_{y}}{\partial z}=-\varepsilon _{0}\varepsilon _{%
\mathrm{rel}}\omega \mathcal{E}_{x},  \label{ME} \\
\mathcal{E}_{z}=\frac{i}{\varepsilon _{0}\varepsilon _{\mathrm{rel}}\omega }%
\frac{\partial \mathcal{H}_{y}}{\partial x},  \notag
\end{gather}%
where $\omega $ is the frequency of the monochromatic carrier, $\varepsilon
_{0}$ and $\mu _{0}$ are the vacuum permittivity and permeability, and $%
\varepsilon _{\mathrm{rel}}=\varepsilon _{\mathrm{bg}}+\varepsilon ^{\mathrm{%
re}}(x)+i\varepsilon ^{\mathrm{im}}(x)$ is the complex relative permittivity
of the $\mathcal{PT}$-symmetric structure, with $x$-dependent real and
imaginary parts, added to the background permittivity, $\varepsilon _{%
\mathrm{bg}}$.

Two different modulation patterns were considered in Ref. \cite{sub},
corresponding, respectively, to a single waveguiding channel or a periodic
guiding structure in the $\left( x,z\right) $ plane. In this article, we
focus on solitary (localized) modes, therefore only the former pattern is
explicitly considered. It is defined by the following transverse ($x$%
-dependent) profile:%
\begin{equation}
\varepsilon _{\mathrm{rel}}(x)=\varepsilon _{\mathrm{bg}}+\mathrm{sech}%
^{2}\left( \frac{x}{d}\right) \left[ p~+i\alpha \mathrm{~sinh}\left( \frac{x%
}{d}\right) \right] ,  \label{channel}
\end{equation}%
where $d$ and $p>0$ represent, severally, the width and depth of the guiding
channel, while $\alpha >0$ is the strength of the gain-loss term. In
accordance with the the general definition of the $\mathcal{PT}$ symmetry,
the real and imaginary parts of the profile are even and odd functions of $x$%
, respectively, cf. Eq. (\ref{minus}).

Eigenmodes for subwavelength beams with propagation constant $b$ are looked
for as solutions to Eq. (\ref{ME}) in the form of%
\begin{equation}
\left\{ \mathcal{E}_{x}(x,z),\mathcal{H}_{y}\left( x,z\right) ,\mathcal{E}%
_{z}\left( x,z\right) \right\} =e^{ibz}\left\{ E_{x}(x)H_{y}\left( x\right)
,E_{z}\left( x\right) \right\} .  \label{eigen}
\end{equation}%
Numerical solution of Eq. (\ref{ME}) with modulation profile (\ref{channel})
has produced three types of the solutions \cite{sub}: (i) ones with real $b>%
\sqrt{\varepsilon _{\mathrm{bg}}}$ represent stable $\mathcal{PT}$-symmetric
beams guided by the channel; (ii) solutions with a complex propagation
constant, which has Re$\left( b\right) >\sqrt{\varepsilon _{\mathrm{bg}}}$,
Im$(b)\neq 0$ represent, as it follows from Eq. (\ref{eigen}), exponentially
growing (unstable) channel-guided modes with broken $\mathcal{PT}$ symmetry,
and (iii) delocalized modes, which are not actually guided by the channel,
have Re$\left( b\right) <\sqrt{\varepsilon _{\mathrm{bg}}}$.

\begin{figure}[tbp]
\centering
\includegraphics[width=0.5\textwidth]{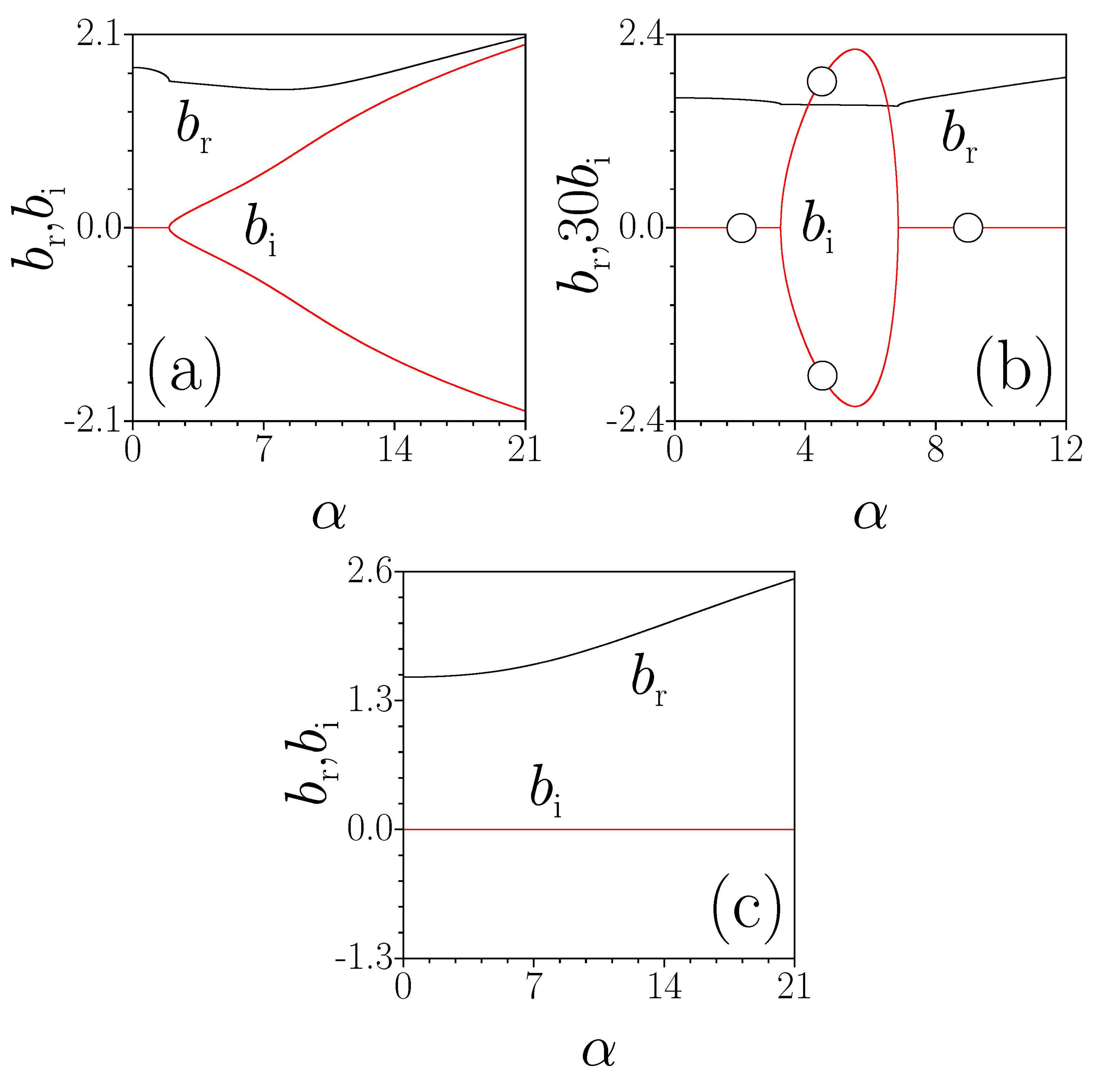}
\caption{Real and imaginary parts of the propagation constant, $b_{\mathrm{r}%
}$ and $b_{\mathrm{i}}$, versus the gain-loss strength, $\protect\alpha $,
in Eq. (\protect\ref{channel}), for the guiding channel with depth $p=1.7$,
and widths $d=120$ nm (a), $d=60$ nm (b), and $d=30$ nm (c) (as per Ref.
\protect\cite{sub}). Circles in panel (b) designate examples of the
eigenmodes displayed in Fig. \protect\ref{fig2}. The underlying wavelength
is $\protect\lambda =632.8$ nm, and the background dielectric permeability
is $\protect\varepsilon _{\mathrm{bg}}=2.25$. The emergence of $b_{\mathrm{i}%
}$ in panels (a) and (b) signals the breaking of the $\mathcal{PT}$
symmetry, while the disappearance of $b_{\mathrm{i}}$ in (b) implies the%
\emph{\ restoration} of the symmetry. In the case shown in (c), the $%
\mathcal{PT}$ symmetry is \emph{never broken}.}
\label{fig1}
\end{figure}

The situation which occurs in a majority of previously studied models is
that, with the increase of the gain-loss strength, $\alpha $, the $\mathcal{%
PT}$ symmetry of the guided states suffers breaking at a critical value, $%
\alpha _{\mathrm{cr}}$. This is indeed observed in the present case in the
nearly-paraxial regime, namely, at $d/\lambda \gtrsim 1/5$, where $\lambda $
is the underlying wavelength of the optical beam (below, following Ref. \cite%
{sub}, particular results are displayed for $\lambda =632.8$ nm (visible
red), and $\varepsilon _{\mathrm{bg}}=2.25$). In particular, at $d=120$ nm,
the breaking of the $\mathcal{PT}$ symmetry takes place at $\alpha _{\mathrm{%
cr}}\approx 1.95$, see Fig. \ref{fig1}(a) (in Fig. 1, the $\mathcal{PT}$
symmetric modes exist at a single value of the propagation constant, as the
underlying wavelength is fixed). However, in the deeply subwavelength
situation, corresponding to essentially smaller channel's widths, such as $%
d=60$ nm $\simeq \lambda /10$ and $30$ nm $\simeq \lambda /20$ (see Figs. %
\ref{fig1}(b,c)), a drastically different situation is observed: in the
former case, the breaking of the\ $\mathcal{PT}$ symmetry is followed its
\emph{restoration} at still larger values of $\alpha $, and in the latter
case the breaking \emph{does not happen} at all.

It is relevant to mention that a similar effect of the spontaneous
restoration of the $\mathcal{PT}$ symmetry, although not the full
elimination of the symmetry breaking, was reported too in some other models
(based on the paraxial, rather than subwavelength, equations), including a
linear discrete system of the Aubry-Andr\'{e} type \cite{Joglekar}, and a
nonlinear model based on the NLSE in 1D \cite{Segev}. Examples of
unbreakable $\mathcal{PT}$ symmetry are known too in simple models with few
degrees of freedom, such as a $\mathcal{PT}$ dimer \cite{Barash2}.

\begin{figure}[tbp]
\centering\includegraphics[width=0.5\textwidth]{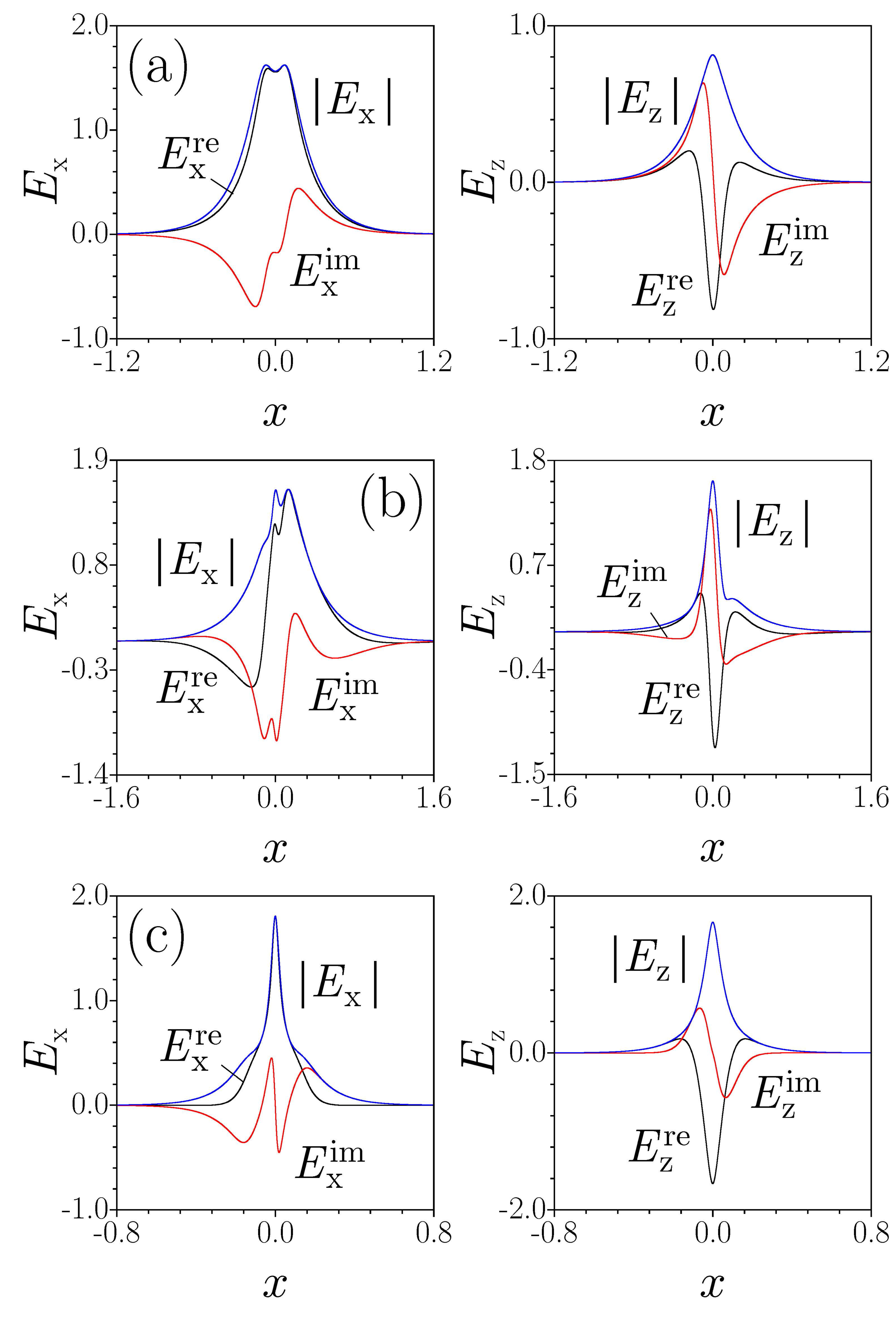}
\caption{Profiles of the guided modes designated by circles in Fig. \protect
\ref{fig1}, at $\protect\alpha =2.0$ (a), $4.5$ (b), and $9.5$ (c), which
are typical modes with unbroken, broken, and \emph{restored} $\mathcal{PT}$
symmetry, respectively (as per Ref. \protect\cite{sub}). The fields are
plotted in dimensionless units, while transverse coordinate $x$ is measured
in $\mathrm{\protect\mu }$m.}
\label{fig2}
\end{figure}

A set of typical eigenmodes of the electromagnetic fields, which correspond,
respectively, to the unbroken, broken, and restored $\mathcal{PT}$ symmetry,
are displayed in Fig. \ref{fig2}. It is clearly seen that, in the case of
the unbroken and restored symmetry, each field component is either spatially
even or odd, while the modal spatial (anti)symmetry is broken too when the $%
\mathcal{PT}$ symmetry does not hold.

\begin{figure}[tbp]
\centering\includegraphics[width=0.5\textwidth]{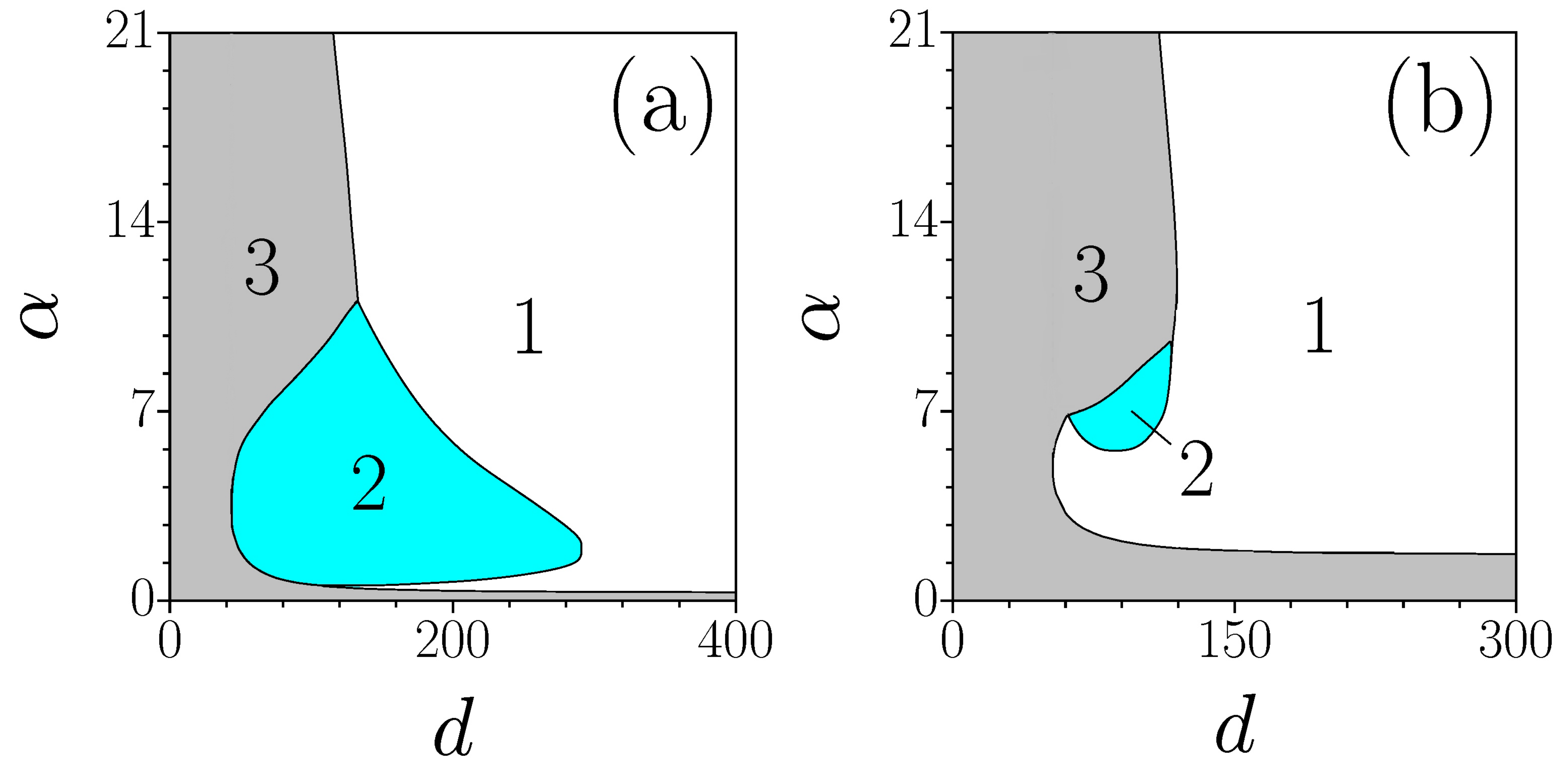}
\caption{Domains of the existence and stability of the $\mathcal{PT}$%
-symmetric modes guided by channel (\protect\ref{channel}), in the plane of
the channel's width, $d$, and gain-loss coefficient, $\protect\alpha $ (as
per Ref. \protect\cite{sub}). The depth of the channel is $p=0.3$ in (a),
which represents a shallow conduit, and $p=1.7$ in (b), representing a deep
one. The symmetry is broken in region \textbf{1}, and unbroken \emph{or
restored} in region \textbf{3}, respectively, while region \textbf{2} does
not support any localized mode.}
\label{fig3}
\end{figure}

Finally, the results of the consideration of the model are summarized in
Fig. \ref{fig3}, which shows regions of the unbroken, broken, and restored $%
\mathcal{PT}$ symmetry in the plane of the essential control parameters,
\textit{viz}., the gain-loss coefficient, $\alpha $, and the width of the
guiding channel, $d$. Relatively small areas where no guided modes exist (in
the latter case, the optical beam coupled into the channel waveguide suffers
delocalization, spreading out into the entire $\left( x,z\right) $ plane)
are shown too. The conclusion suggested by Fig. \ref{fig3} is quite clear:
in the near-paraxial regime, corresponding to a relatively broad guiding
channel, with $d\gtrsim 120$ nm, the usual scenario of the $\mathcal{PT}$%
-symmetry breaking, following the increase of $\alpha $, is observed.
However, in the deeply subwavelength region, the symmetry (hence, the
stability of the guided modes too) is either readily \emph{restored} with
the further increase of $\alpha $, or is \emph{never broken}. Figure \ref%
{fig3} demonstrates too that region \textbf{3} of the unbroken and restored
stability tends to expand, although not very dramatically, with the increase
of the channel's depth, $p$ (see Eq. \ref{channel})), while, quite
naturally, the delocalization area \textbf{2} shrinks.

\section{Unbreakable $\mathcal{PT}$-symmetric solitons in one dimension}

The 1D model which is capable to support solitons with unbreakable $\mathcal{%
PT}$ symmetry by means of the self-defocusing nonlinearity with the local
strength, $S(\eta )$, growing from the center to infinity, as a function of
coordinate $\eta $, is based on the NLSE for the amplitude of the
electromagnetic field, $u$ \cite{unbreakable}:%
\begin{equation}
i\frac{\partial u}{\partial \xi }+\frac{1}{2}\frac{\partial ^{2}u}{\partial
\eta ^{2}}-S(\eta )|u|^{2}u=-iR(\eta )u,  \label{qq}
\end{equation}%
where $\xi $ is the propagation coordinate, and $S(\eta )$ provides for the
self-trapping of 1D solitons under that condition that$\ S(\eta )$ grows
faster than $|\eta |$ at $|\eta |\rightarrow \infty $ \cite%
{Barcelona1,Barcelona2}. Here, following Ref. \cite{unbreakable}, we adopt a
steep anti-Gaussian modulation profile,%
\begin{equation}
S(\eta )=\left( 1+\sigma \eta ^{2}\right) \exp \left( \frac{1}{2}\eta
^{2}\right) ,  \label{Gauss}
\end{equation}%
where coefficients equal to $1$ and $1/2$ may be fixed to these values by
means of rescaling of a more general expression. Further, the spatially-odd
gain-loss modulation profile is adopted also as it was done in Ref. \cite%
{unbreakable}:%
\begin{equation}
R(\eta )=\beta \eta \exp \left( -\Gamma \eta ^{2}\right) ,  \label{gamma}
\end{equation}%
with $\beta >0$ and $\Gamma \geq 0$.

An advantage of fixing the profiles in the form of Eqs. (\ref{Gauss}) and (%
\ref{gamma}) is that they admit a particular exact solution for the
self-trapped $\mathcal{PT}$-symmetric soliton \cite{unbreakable}, provided
that $\Gamma =0$ is set in Eq. (\ref{gamma}):%
\begin{equation}
u\left( \eta ,\xi \right) =\frac{1}{2\sqrt{2\sigma }}\exp \left( ib\xi
-2i\beta \eta -\frac{1}{4}\eta ^{2}\right) ,  \label{exact}
\end{equation}%
at a single value of the propagation constant:%
\begin{equation}
b=-\left( 2\beta ^{2}+\frac{1}{4}+\frac{1}{8\sigma }\right) .  \label{b}
\end{equation}%
The availability of the exact solution is principally important for
establishing the concept of the \textit{unbreakability} of the $\mathcal{PT}$
symmetry: obviously, the solution given by Eqs. (\ref{exact}) and (\ref{b})
exist for \emph{arbitrarily large} values of the gain-loss strength, $\beta $%
, there being no critical value beyond which solitons would not exist.
Moreover, in Ref. \cite{unbreakable} it was checked, at least in a part of
the parameter plane $\left( \beta ,\sigma \right) $, that the exact solitons
are stable.

It is relevant to stress that the model with the sufficiently quickly
growing nonlinearity coefficient $S(\eta )$ is \emph{nonlinearizable}: the
form of decaying tails of generic self-trapped modes can be investigated
analytically (it turns out to be the same as in the particular exact
solution (\ref{exact})), but it is necessary to keep the nonlinear term in
Eq. (\ref{qq}) for this purpose \cite{Barcelona0,Barcelona1}. Accordingly,
the linear spectrum of the present model cannot be defined, the respective
concept of the $\mathcal{PT}$ symmetry and its breaking or unbreakability
being a nonlinear one too. The same pertains to the 2D model considered in
the next section.

Numerical solution of Eq. (\ref{qq}) produces many families of complex
solitons with real propagation constant $b$, in the form of%
\begin{equation}
u\left( \eta ,\xi \right) =\exp \left( ib\xi \right) \left[ w_{\mathrm{r}%
}(\eta )+iw_{\mathrm{i}}(\eta )\right] ,  \label{w}
\end{equation}%
which may be naturally identified as fundamental solitons, dipoles,
tripoles, quadrupoles, and so on. These solution types feature profiles of $%
|w(\eta )|\equiv \sqrt{w_{\mathrm{r}}^{2}(\eta )+w_{\mathrm{i}}^{2}(\eta )}$
with, respectively, one, two, three, etc. peaks (local maxima). Solitons are
characterized by their integral power,%
\begin{equation}
U=\int_{-\infty }^{+\infty }\left\vert w(\eta )\right\vert ^{2}d\eta .
\label{U}
\end{equation}

\begin{figure}[tbp]
\centering\includegraphics[width=0.5\textwidth]{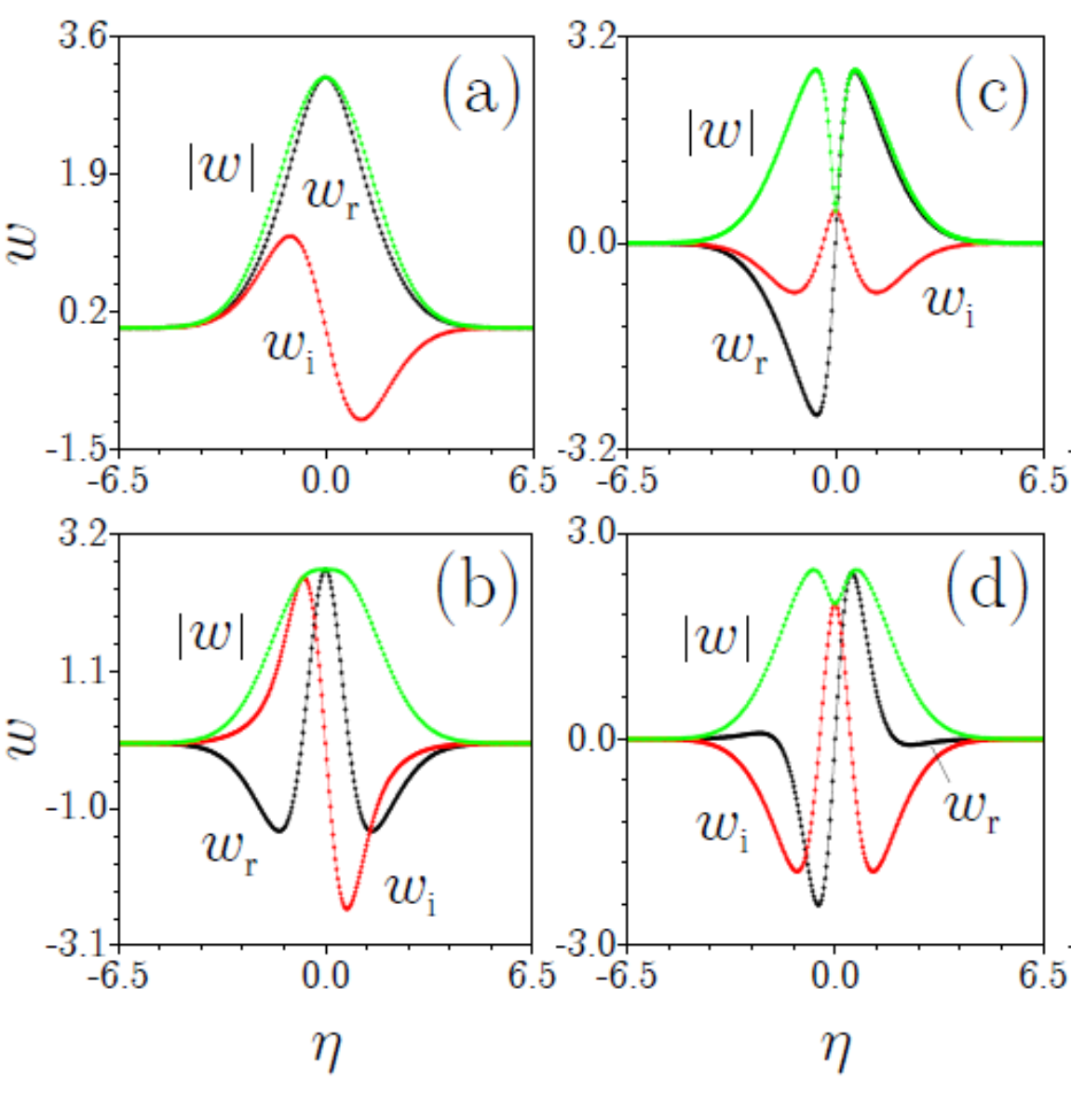}
\caption{Profiles of fundamental (a),(b) and dipole (c),(d) stable
one-dimensional solitons, found as numerical solutions of Eq. (\protect\ref%
{qq}) with $\protect\sigma =0$ and $\protect\gamma =1/2$, for a fixed value
of the propagation constant, $b=-10$ (as per Ref. \protect\cite{unbreakable}%
). Panels (a), (c) and (b), (d) pertain, severally, to $\protect\beta =1.04$
and $3.47$.}
\label{fig4}
\end{figure}

\begin{figure}[tbp]
\centering\includegraphics[width=0.5\textwidth]{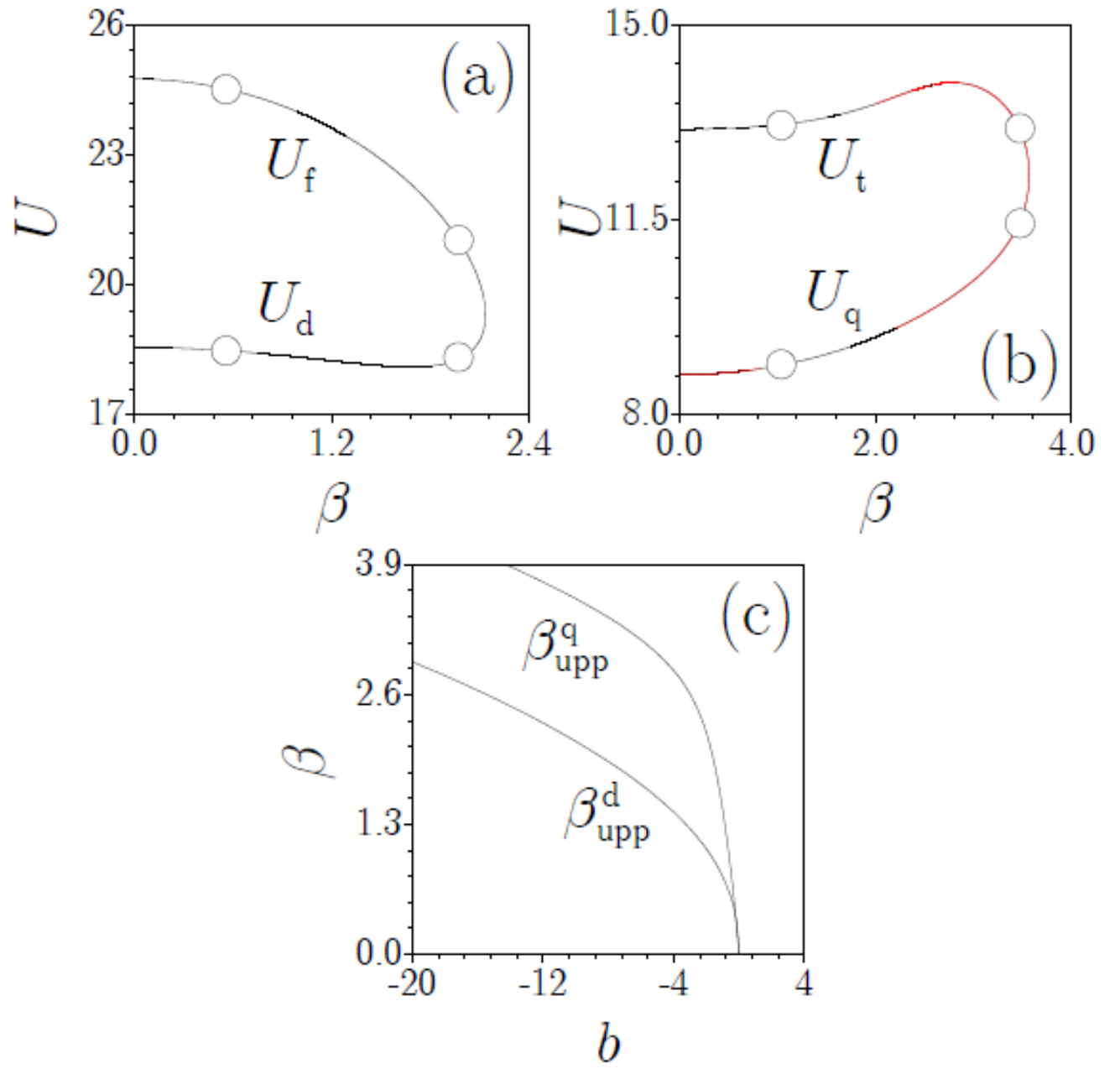}
\caption{The solitons' integral power, defined in Eq. (\protect\ref{U}), vs.
the gain-loss strength, $\protect\beta $, for branches of the fundamental
and dipole solitons (a), and ones of the tripole and quadrupole types (b)
(as per Ref. \protect\cite{unbreakable}). In (b), black and red segments
designate stable and unstable solitons, respectively (the fundamental and
dipole solitons are completely stable in their existence areas). Circles in
(a) correspond to examples of the solutions shown in Fig. \protect\ref{fig4}
(circles in (b) correspond to examples of stable and unstable tripole and
quadrupole solitons which can be found in Ref. \protect\cite{unbreakable},
but are not shown here). The families are produced for $\protect\sigma =0$
in Eq. (\protect\ref{Gauss}), $\protect\gamma =1/2$ in Eq. (\protect\ref%
{gamma}), and fixed propagation constant, $b=-10$. The fundamental and
dipole families merge at $\protect\beta \approx 2.135$, while the tripole
and quadrupole ones merge at $\protect\beta \approx 3.565$. (c) The critical
(\textquotedblleft upper") value, $\protect\beta _{\mathrm{upp}}^{\mathrm{d}%
} $, at which the fundamental and dipole branches merge, vs. the propagation
constant, $b$. Curve $\protect\beta _{\mathrm{upp}}^{\mathrm{q}}(b)$ shows
the same for the merger of the tripole and quadrupole branches.}
\label{fig5}
\end{figure}

Characteristic examples of stable fundamental and dipole solutions are
displayed in Fig. \ref{fig4} (they were obtained for $\sigma =0$, in which
case exact soliton (\ref{exact}) does not exist, but numerically found
solitons are available and may be stable). It is seen that the increase of
the gain-loss coefficient, $\beta $, makes the shape of the solitons more
complex, but the fundamental and dipole solitons remain fully stable as long
as they exist, while higher-order tripoles and quadrupoles have \ both
stability and instability areas \cite{unbreakable} (as briefly shown in Fig. %
\ref{fig5}(b)).

Most essential results characterizing the behavior of solitons in the
present model are collected in Fig. \ref{fig5}. In particular, Fig. \ref%
{fig5}(a) shows that, at fixed $b$, branches of the fundamental and dipole
solitons, remaining completely stable, merge and disappear, with the
increase of the gain-loss coefficient, $\beta $, at a critical
(\textquotedblleft upper") value, which is $\beta _{\mathrm{upp}}\approx
2.135$ in Fig. \ref{fig5}(a). However, stable fundamental and dipole soliton
can be found at arbitrarily high values of $\beta $, as demonstrated by the
lower curve in Fig. \ref{fig5}(c), which shows the critical value $\beta _{%
\mathrm{upp}}$ vs. $b$: obviously, $\beta $ may become indefinitely large
with the increase of $|b|$. In addition, the upper curve shows the growth
with $|b|$ of a similar critical (\textquotedblleft upper") value at which
another pair of solitons, \textit{viz}., tripoles and quadrupoles, merge, as
can be seen in Fig. \ref{fig5}(b) (however, unlike the fundamental and
dipole modes, the tripole and quadrupole branches become unstable prior to
the merger, as seen in \ref{fig5}(b)).

\section{Unbreakable $\mathcal{PT}$-symmetric solitons in two dimensions}

\subsection{The model and analytical solutions}

Results presented in the above sections summarize findings originally
published in Refs. \cite{sub} and \cite{unbreakable}, respectively. Here we
report previously unpublished analytical and numerical results obtained for
2D generalizations of the model based on Eq. (\ref{qq}). The 2D model with
transverse coordinates $\left( x,y\right) $ and propagation distance $z$ is
based on the following NLSE for the amplitude of the electromagnetic field, $%
w\left( x,y,z\right) $:

\begin{equation}
i\frac{\partial w}{\partial z}+\frac{1}{2}\left( \frac{\partial ^{2}w}{%
\partial x^{2}}+\frac{\partial ^{2}w}{\partial y^{2}}\right) - S
(r)|w|^{2}w=i R \left( x,y\right) w,  \label{NLS}
\end{equation}%
where $r\equiv \sqrt{x^{2}+y^{2}}$ is the radial coordinate, and the
nonlinearity-modulation profile is chosen similar to its 1D counterpart (\ref%
{Gauss}):
\begin{equation}
S(r)=\left( 1+\sigma r^{2}\right) \exp \left( r^{2}\right) ,  \label{sigma}
\end{equation}%
with $\sigma \geq 0$.

Here we consider two different versions of the gain-loss spatial profile: a
quasi-1D one, symmetric only with respect to $x$:%
\begin{equation}
R\left( x,y\right) =\beta _{0}x\exp \left( -\Gamma r^{2}\right) ,  \label{x}
\end{equation}%
and a profile symmetric with respect to $x$ and $y$, which may be called a
fully 2D one:
\begin{equation}
R\left( x,y\right) =\beta _{0}xy\exp \left( -\Gamma r^{2}\right) ,
\label{xy}
\end{equation}%
with constants $\Gamma \geq 0$ and $\beta _{0}>0$.

Stationary solutions with a real propagation constant, $b$, are looked for
as
\begin{equation}
w\left( x,y\right) =\exp \left( ibz\right) W\left( x,y\right) ,  \label{uU}
\end{equation}%
with complex function $W\left( x,y\right) $ satisfying the following
equation:%
\begin{equation}
bW=\frac{1}{2}\left( \frac{\partial ^{2}W}{\partial x^{2}}+\frac{\partial
^{2}W}{\partial y^{2}}\right) - S (r)|W|^{2}W - i R \left( x,y\right) W.
\label{UU}
\end{equation}

In the case of $\Gamma =0$ in Eqs. (\ref{x}) and (\ref{xy}), Eq. (\ref{UU}),
with $\sigma $ and $R\left( x,y\right) $ taken in the form of Eqs. (\ref%
{sigma}) and (\ref{x}), gives rise to an exact analytical solution:%
\begin{equation}
W\left( x,y\right) =W_{0}\exp \left( -\frac{1}{2}r^{2}-i\beta _{0}x\right) ,
\label{exact1}
\end{equation}%
(cf. the 1D solution (\ref{exact})), with%
\begin{equation}
W_{0}^{2}=\frac{1}{2\sigma },~b=-\left( 1+\frac{\beta _{0}^{2}}{2}+\frac{1}{%
2\sigma }\right) .  \label{exact1parameters}
\end{equation}%
This solution exists for all values of the control parameters, $\beta _{0}$
and $\sigma $, except for $\sigma =0$. Further, Eq. (\ref{U}) with $\sigma $
and $R\left( x,y\right) $ taken in the form of Eqs. (\ref{sigma}) and (\ref%
{xy}), where $\Gamma =0$ is again fixed, also gives rise to an exact
solution:%
\begin{equation}
W\left( x,y\right) =W_{0}\exp \left( -\frac{1}{2}r^{2}-\frac{1}{2}i\beta
_{0}xy\right) ,  \label{exact2}
\end{equation}%
this time with%
\begin{equation}
W_{0}^{2}=\frac{1}{2\sigma }\left( 1-\left( \frac{\beta _{0}}{2}\right)
^{2}\right) ,~b=-\left[ 1+\frac{1}{2\sigma }\left( 1-\left( \frac{\beta _{0}%
}{2}\right) ^{2}\right) \right] .  \label{exact2parameters}
\end{equation}%
This solution exists if Eq. (\ref{exact2parameters}) yields $W_{0}^{2}>0$,
i.e., $\beta _{0}<2$ and $\sigma >0$.

Another exact solution of Eq. (\ref{U}), with $\sigma $ and $R\left(
x,y\right) $ again taken in the form of Eqs. (\ref{sigma}) and (\ref{xy}),
exists under the special condition,
\begin{equation}
\beta _{0}=2,\sigma =0,\Gamma =0.  \label{special}
\end{equation}%
This solution is also found in the form of ansatz (\ref{exact2}), precisely
with $\frac{1}{2}\beta _{0}$ replaced by $1$, as per Eq. (\ref{special}).
However, unlike the solution represented by Eqs. (\ref{exact2}) and (\ref%
{exact2parameters}), this time it is not a single one, but a \emph{%
continuous family }of exact solutions, with \emph{arbitrary amplitude} $%
W_{0} $, and propagation constant%
\begin{equation}
b=-\left( 1+W_{0}^{2}\right) .  \label{k}
\end{equation}%
The possibility to obtain the continuous family of the exact 2D solitons,
instead of an isolated one, is a compensation for selecting the special
values of the parameters, as fixed by Eq. (\ref{special}).

The exact solutions clearly suggest that the quasi-1D model, based on Eq. (%
\ref{x}), features the unbreakable $\mathcal{PT}$ symmetry, as the
respective solution, given by Eqs. (\ref{exact1}) and (\ref{exact1parameters}%
), exists for an arbitrarily large strength of the gain-loss term, $\beta
_{0}$. On the other hand, the full 2D model, based on Eq. (\ref{xy}), gives
rise to the exact solutions, in the form of Eqs. (\ref{exact2}), (\ref%
{exact2parameters}) or (\ref{special}), (\ref{k}), which exist only at $%
\beta _{0}\leq 2$, hence the unbreakability of the $\mathcal{PT}$ symmetry
is not guaranteed in the latter case.

\subsection{Numerical results}

\subsubsection{The quasi-1D model}

The exact solution of the model with the quasi-1D gain-loss modulation,
given by Eqs. (\ref{exact1}) and (\ref{exact1parameters}) can be embedded
into a family of solitons produced by a numerical solution of Eq. (\ref{UU}%
), with $S(r)$ and $R\left( x,y\right) $ taken as per Eqs. (\ref{sigma}) and
(\ref{x}), respectively (the latter is taken here with $\Gamma =0$). The
stationary 2D solutions were constructed by means of the Newton's conjugate
gradient method. Then, the stability of the stationary states was identify
by numerical computation of eigenvalues of small perturbations, using
linearized equations for perturbations around the stationary solitons. This
computation was performed with the help of the spectral collocation method.
Finally, the stability prediction, based on the eigenvalues, was verified
through direct simulations of the perturbed evolution of the solitons.

\begin{figure}[tbp]
\centering
\subfigure{\includegraphics[width=0.45\textwidth]{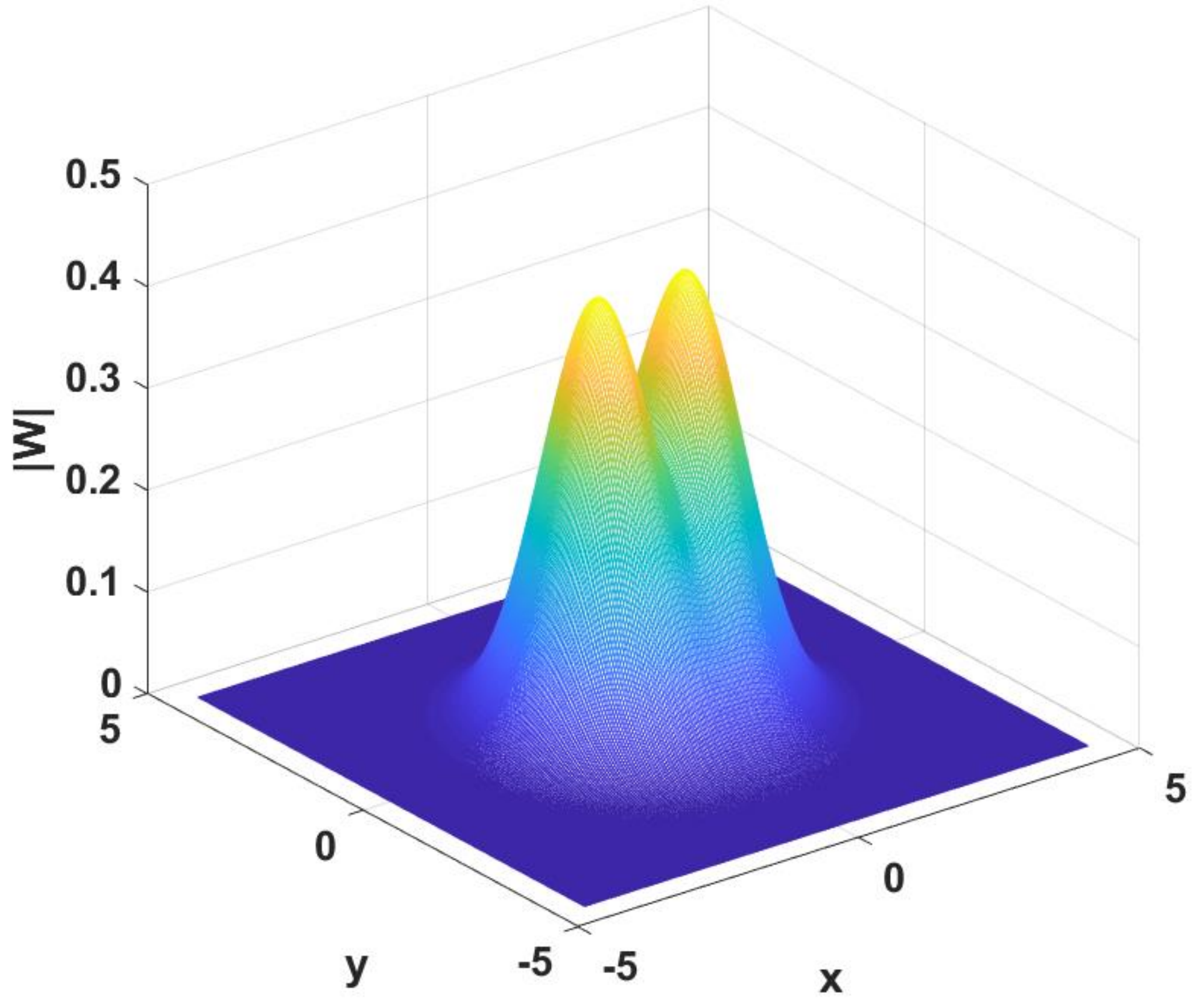}} \newline
\subfigure{\includegraphics[width=0.45\textwidth]{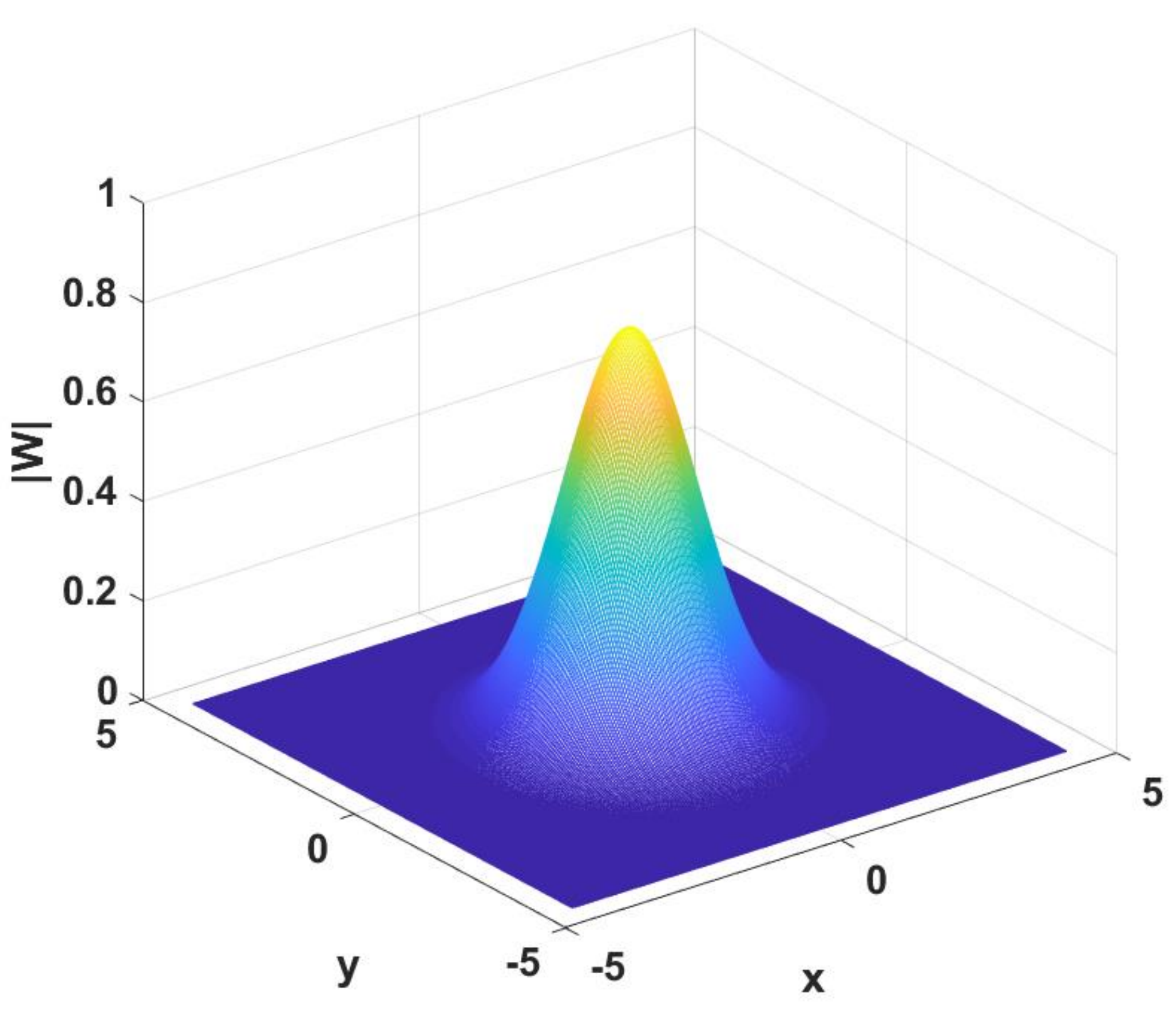}}
\caption{Typical examples of 2D $\mathcal{PT}$-symmetric solitons produced
by the model with the quasi-1D gain-loss profile defined by Eq. (\protect\ref%
{x}). The top and bottom panels display, severally, a stable single-peak
soliton with propagation constant $b=-2$, and an unstable dual-peak one with
$b=-2.7$. In both cases, other parameters are $\protect\beta _{0}=0.8$, $%
\protect\sigma =1$, and $\Gamma =0$.}
\label{fig6}
\end{figure}

Generic examples of numerically found stable and unstable solitons, which
may have single- and dual-peak shapes, are shown in Fig. \ref{fig6}. In
accordance with these examples, all the double-peak solitons are unstable,
and almost all the single-peak ones are stable. In particular, all the exact
solutions, given by Eqs. (\ref{exact1}) and (\ref{exact1parameters}), are
found to be stable.

\begin{figure}[tbp]
\centering\includegraphics[width=0.5\textwidth]{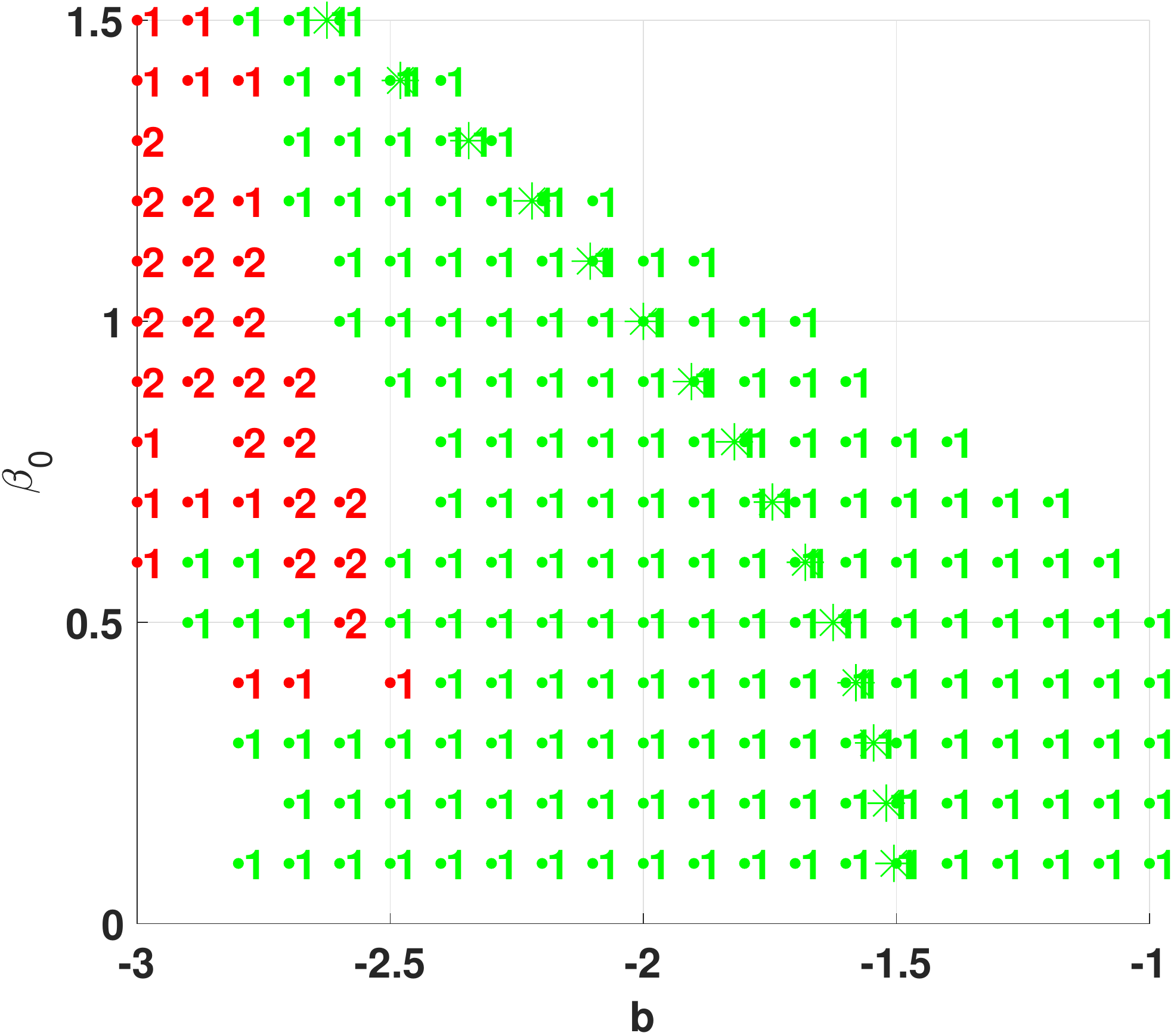}
\caption{(Color online) The stability chart for the solitons supported by
quasi-1D $\mathcal{PT}$-symmetric gain-loss profile (\protect\ref{x}) with $%
\Gamma =0$, in the case of $\protect\sigma =1$ in Eq. (\protect\ref{sigma}).
Exact soliton solutions, given by Eqs. (\protect\ref{exact1}) and (\protect
\ref{exact1parameters}), are indicated by stars (they all are stable), while
stable and unstable numerically found solitons are shown by green and red
dots, respectively. Numbers near the dots denote the number of peaks in each
soliton (one or two). No soliton solutions were found in white areas.}
\label{fig7}
\end{figure}

Results of the stability analysis for the $\mathcal{PT}$-symmetric solitons
in the model with the quasi-1D shape of the gain-loss term, based on the
eigenvalue computation, are summarized by the stability chart in the plane
of $\left( b,\beta _{0}\right) $, i.e., the soliton's propagation constant
and strength of the gain-loss term in Eq. (\ref{x}), which is displayed in
Fig. \ref{fig7}. Direct simulations completely corroborate the predictions
produced by the stability eigenvalues. In particular, the solitons which are
predicted to be unstable get destructed, decaying in the course of the
perturbed evolution. This figure corroborates the unbreakable character of
the $\mathcal{PT}$ symmetry in the model, as the stability region does not
exhibit a boundary at large values of $\beta _{0}$.

\begin{figure}[tbp]
\centering\includegraphics[width=0.5\textwidth]{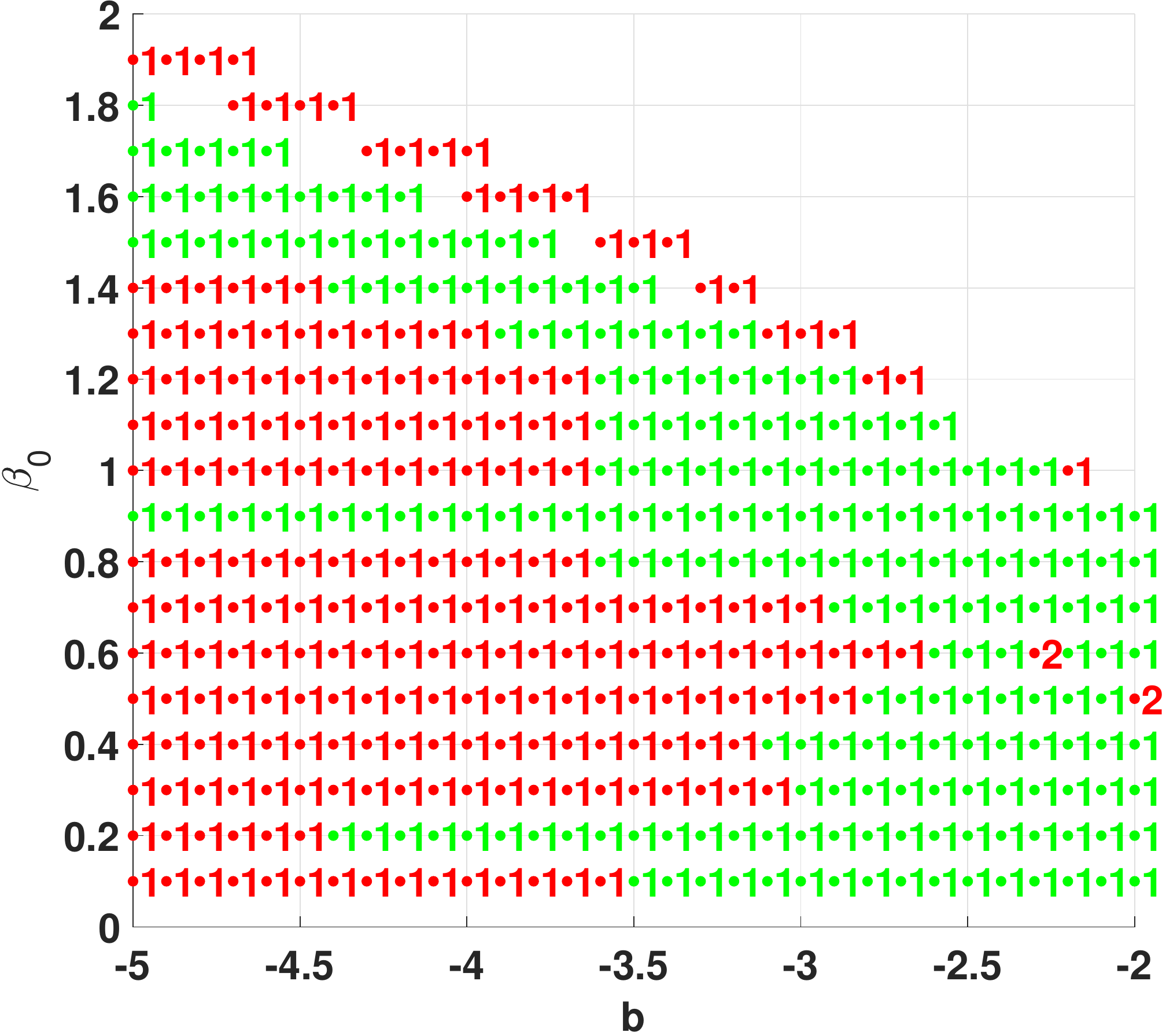}
\caption{The same as in Fig. \protect\ref{fig7} (the stability chart for $%
\mathcal{PT}$-symmetric solitons), but for $\protect\sigma =0$ in Eq. (%
\protect\ref{sigma}). Note an essential reduction of the stability area in
comparison with its counterpart in Fig. \protect\ref{fig7}.}
\label{fig8}
\end{figure}

The stability chart, drawn in Fig. \ref{fig7} for $\sigma =1$ in Eq. (\ref%
{sigma}), is quite similar to its counterparts produced at other values of $%
\sigma >0$. The situation is different in the case of $\sigma =0$, when the
exact solution given by Eqs. (\ref{exact1}) and (\ref{exact1parameters})
does not exist. The respective stability chart, displayed in Fig. \ref{fig8}%
, demonstrates essential differences from the one in Fig. \ref{fig7}: the
stability area is conspicuously smaller, all unstable solutions, as well as
stable ones, featuring the single-peak shape.

\subsubsection{The full 2D model}

\begin{figure}[tbp]
\centering
\subfigure{\includegraphics[width=0.5\textwidth]{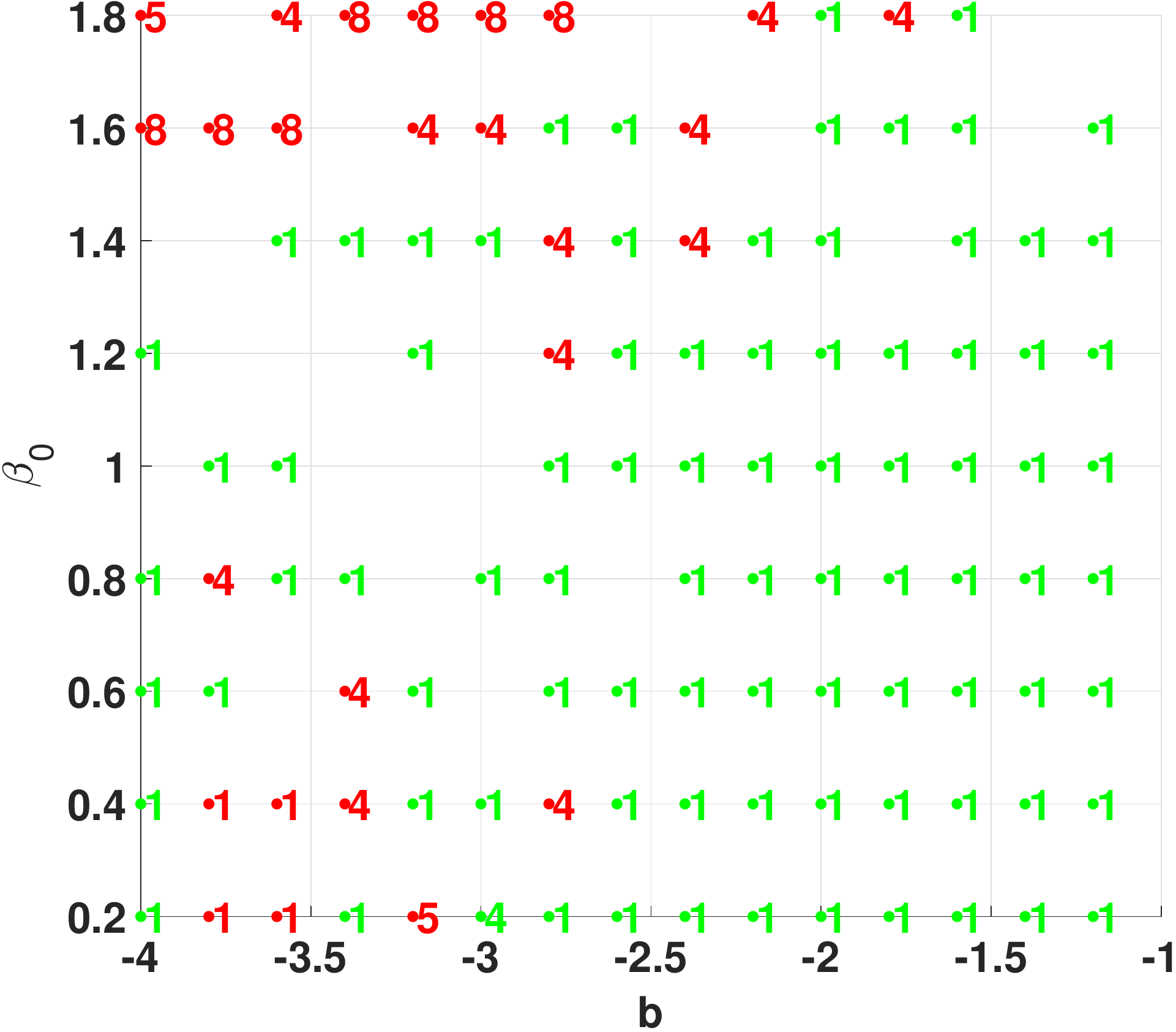}} \newline
\subfigure{\includegraphics[width=0.5\textwidth]{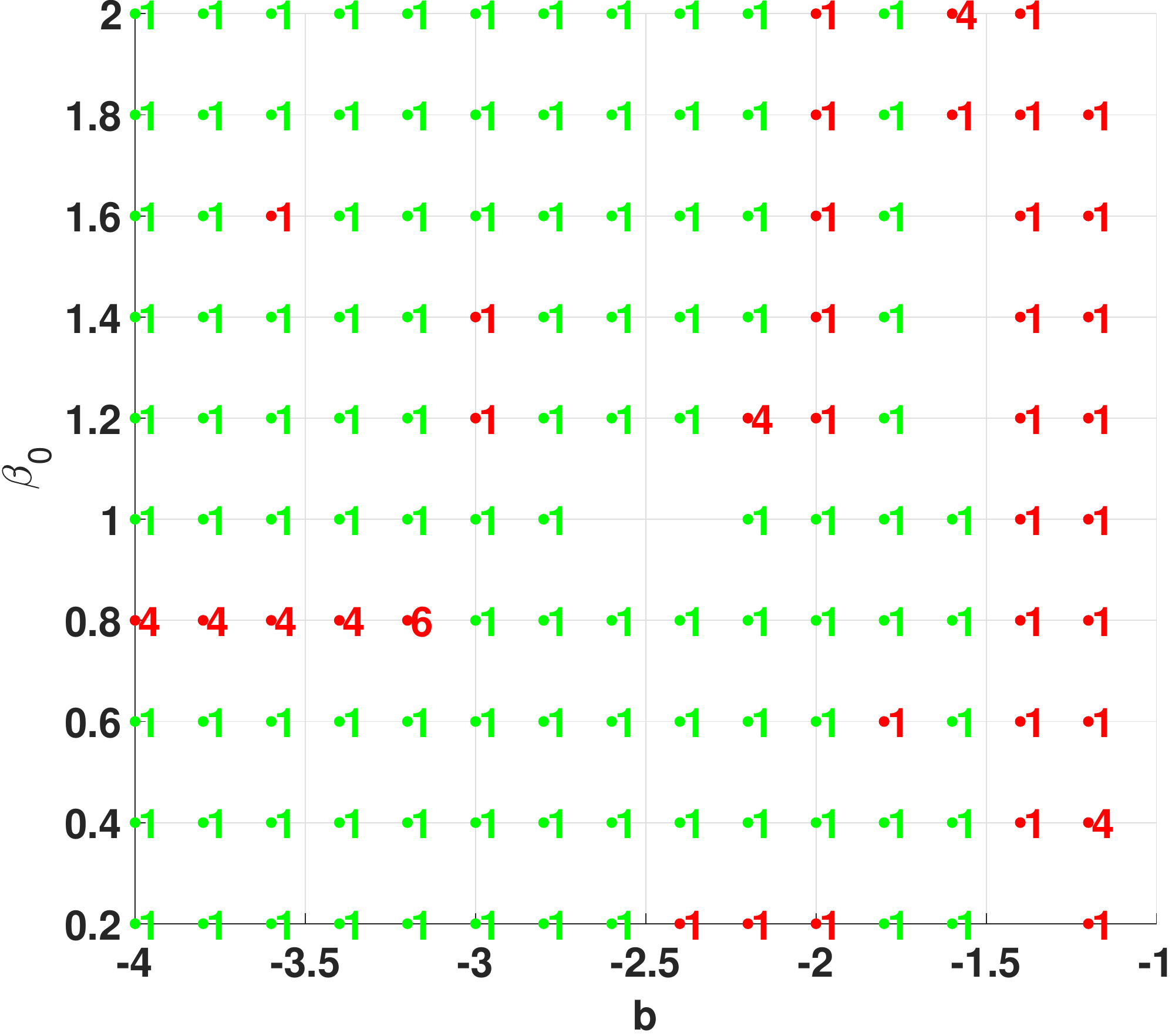}}
\caption{The same (stability charts) as in Figs. \protect\ref{fig7} and
\protect\ref{fig8}, but for the full 2D model based on Eq. (\protect\ref{xy}%
), with $\Gamma =0.5$ , $\protect\sigma =1$ and $\protect\sigma =0$ in the
top and bottom panels, respectively.}
\label{fig9}
\end{figure}

A drastic difference produced by the stability analysis for exact solutions
of the full 2D model, given by Eqs. (\ref{exact2}) and (\ref%
{exact2parameters}) for $\sigma >0$, $\Gamma =0$ and arbitrary $\beta _{0}$,
and by Eq. (\ref{k}) for the special case (\ref{special}), is that these
solutions are completely \emph{unstable}. Furthermore, all numerical
solutions found in the full 2D model with $\Gamma =0$ in Eq. (\ref{xy}) are
unstable too. The stabilization in this model may be provided by $\Gamma >0$%
, i.e., by confining the spatial growth of the local gain and loss in Eq. (%
\ref{xy}). For fixed $\sigma $, there is a minimum value $\Gamma _{\min }$
of $\Gamma $ which provides for the stabilization. In fact, $\Gamma _{\min }$
depends on the size of the solution domain: in an extremely large domain,
one may find very broad solitons, i.e., ones with very small $b$ (see Eq. (%
\ref{UU})), at any $\Gamma >0$. Practically speaking, the size of the domain
is always finite, as the steep growth of $S\left( r\right) $, defined as per
Eq. (\ref{sigma}), cannot extend to infinity. As shown in Refs. \cite%
{Barcelona0}-\cite{Barcelona10}, it is sufficient to secure the adopted
modulation profile of $S(r)$ on a scale which is essentially larger than a
characteristic size of the soliton supported by this profile. Thus, we have
concluded that, for instance, in the domain of size $|x|,|y|~\leq 9$ the
solitons are stable in the model with $\sigma =1$ in Eq. (\ref{sigma}) at $%
\Gamma \geq 0.2$ in Eq. (\ref{xy}), being explicitly unstable, e.g., at $%
\Gamma =0.1$. Typical examples of the stability charts for the $\mathcal{PT}$%
-symmetric solitons, numerically produced in the full 2D model with $\beta
>0 $, are displayed in Fig. \ref{fig9}. Naturally, the stability area
expands with the increase of $\Gamma $. It is worthy to note that Fig. \ref%
{fig9}(b) clearly suggests that the $\mathcal{PT}$ symmetry in the model
with the full 2D modulation of the gain-loss term may also be unbreakable,
as the stability chart does features no upper boundary.

\begin{figure}[tbp]
\centering\includegraphics[width=0.45\textwidth]{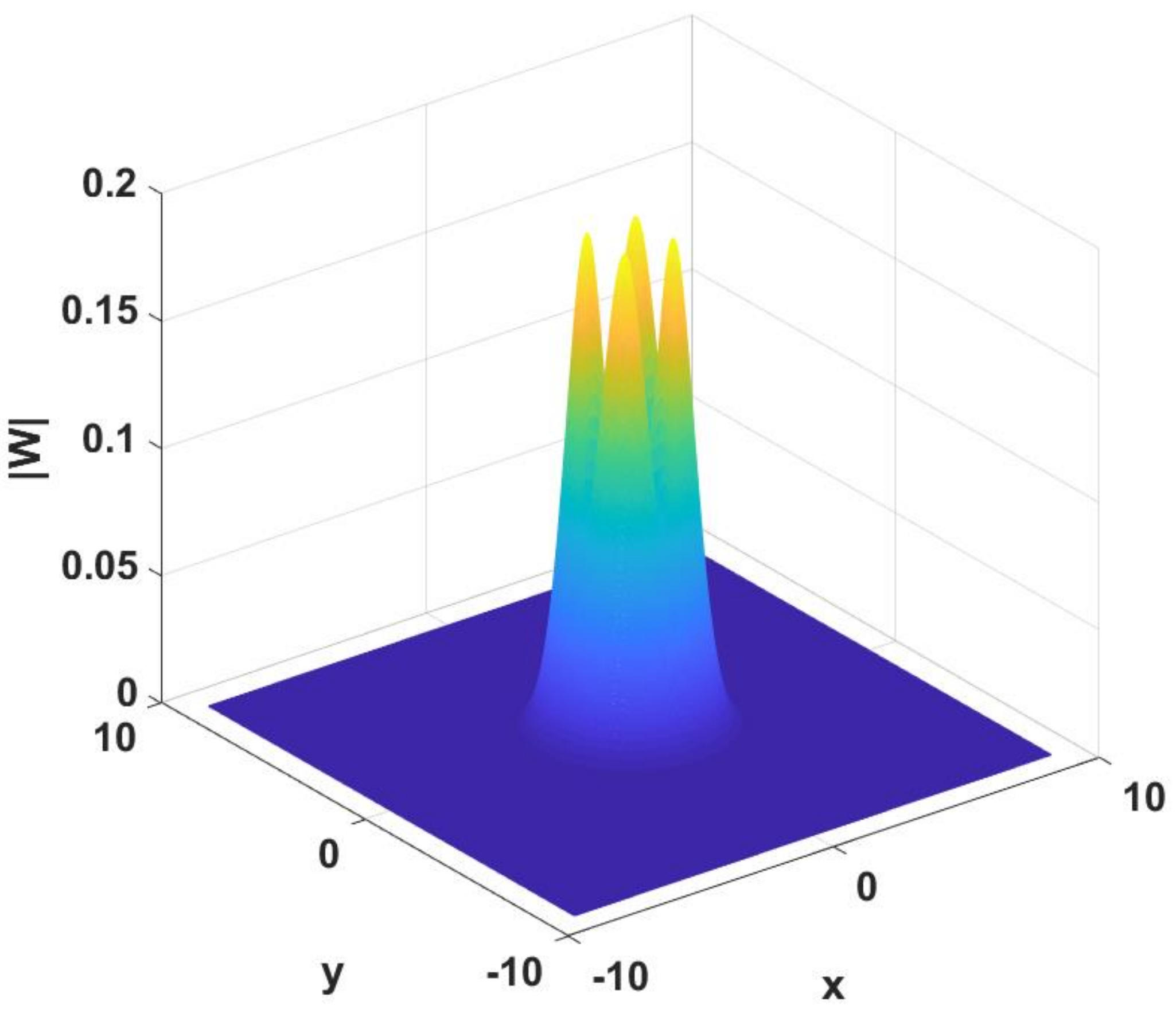}
\caption{An example of a stable $\mathcal{PT}$-symmetric four-peak soliton
with propagation constant $b=-3$, found in the full 2D models with $\protect%
\sigma =1$ in Eq. (\protect\ref{sigma}) and $\Gamma =0.5$ in Eq. (\protect
\ref{xy}).}
\label{fig10}
\end{figure}

These charts include unstable and (very few) stable solitons with multi-peak
shapes. Indeed, taking larger $\Gamma $, i.e., stronger confinement of the
gain and loss in Eq. (\ref{xy}), it is possible to find \emph{stable}
multi-peak solitons with rather complex shapes, an example being a stable
four-peak soliton displayed in Fig. \ref{fig10} for $\Gamma =0.5$.

The results for the quasi-1D and full 2D systems, reported in this section,
do not provide an exhaustive analysis of these models. A comprehensive
analysis, including, in particular, the consideration of possible solitons
with embedded vorticity, will be presented elsewhere.

\section{Conclusion}

The objective of this article is to summarize theoretical results which
demonstrate the stabilization of the $\mathcal{PT}$ symmetry in both linear
and nonlinear systems, making it possible to produce $\mathcal{PT}$%
-symmetric states at arbitrarily large values of the strength of the
gain-loss terms in the system, i.e., of the coefficient in front of the
non-Hermitian part of the underlying $\mathcal{PT}$-symmetric Hamiltonian.
In Sections II and III, we have surveyed previously reported results
obtained in two altogether different settings. Namely, the possibility of
the restoration and complete stabilization of the $\mathcal{PT}$ symmetry in
the linear nanophotonic model of the waveguiding channel with a
subwavelength width, the analysis of which is based on the full system of
the Maxwell's equations, was recapitulated in Section II. The full
stabilization, i.e., removal of the symmetry-breaking transition, takes
place in the deeply subwavelength region. In Section III we have summarized
results concerning the possibility of finding stable 1D solitons supported
by the model with arbitrarily large values of the gain-loss coefficient,
where the self-trapping of the solitons is provided by the self-defocusing
nonlinearity with the local strength growing fast enough from the center to
periphery.\ The model admits a particular exact solution for the fundamental
soliton, the families of both fundamental and dipole modes being entirely
stable.

Section IV has presented new results for the unbreakable $\mathcal{PT}$%
-symmetric solitons in two 2D extensions of the 1D model, \textit{viz}.,
with the quasi-1D and full 2D modulation profiles of the local gain-loss
coefficient. These models also admit particular exact solutions, this time
for 2D solitons. As a result, it is found the quasi-1D model readily gives
rise to the stable family of fundamental (single-peak) 2D solitons for an
arbitrarily large strength of the gain-loss term, while dual-peak ones are
unstable. On the other hand, the stability of the solitons in the model with
the full 2D $\mathcal{PT}$ symmetry requires to impose spatial confinement
on the gain-loss term. Further results for the 2D models will be presented
elsewhere.

\section*{Acknowledgments}

This work was supported, in part, by Grant No. 2015616 from the joint
program in physics between the NSF and Binational (US-Israel) Science
Foundation, and by Grant No. 1286/17 from the Israel Science Foundation.

\end{document}